ARTICLE

# First-principles high-throughput screening of ruthenium compounds for advanced interconnects

Gyungho Maeng,[a] Subeen Lim,[a] Bonggeun Shong,[b] and Yeonghun Lee*[ac]

As interconnect dimensions continue to shrink, the industry-standard copper faces a critical increase in resistivity, presenting a significant hurdle to overall device performance. To overcome this limitation, this work investigates the potential of ruthenium (Ru)-based compounds, encompassing binary, ternary, and quaternary systems, as viable alternatives to copper (Cu). Ruthenium is regarded as a strong candidate, owing to its inherent advantages in reliability and more favorable resistivity scaling at reduced dimensions. Moreover, forming compounds offers an effective strategy to engineer novel properties, expanding the material design space beyond the constraints of pure metals. Utilizing a high-throughput screening methodology, we systematically investigated a broad spectrum of 2,106 Ru-based compounds to identify candidates with superior electronic transport and reliability characteristics. Consequently, we successfully identified a total of 61 promising candidates that exhibit excellent resistivity scaling behavior and enhanced reliability. These findings demonstrate that Ru-based compounds offer a viable pathway to overcome the scaling limitations of next-generation interconnects.

## 1 Introduction

Following the transition from aluminum, copper (Cu) has served as the long-standing industry-standard material for interconnects. However, as device dimensions scale down to the sub-nanometer regime, the industry faces a critical challenge: a dramatic increase in electrical resistivity, leading to significant resistance-capacitance (RC) delays that bottleneck overall device performance. This resistivity size effect is primarily driven by increased electron scattering at surfaces and grain boundaries phenomena, which become dominant when the feature size falls below the material's electron mean free path (MFP) [1–4]. Fundamentally, this limitation stems from the relatively long MFP of Cu, approximately 39 nm [5], which renders it highly sensitive to dimensional scaling.

In this context, ruthenium (Ru) has emerged as a compelling alternative candidate [6–9]. With a significantly shorter MFP of approximately 6.7 nm, Ru exhibits much lower sensitivity to scaling effects compared to Cu [6,10]. Furthermore, the practical implementation of Cu requires thick, non-scalable diffusion barrier and liner layers, which consume a growing fraction of the interconnect volume and diminish the conductive cross-section [11,12]. Ru directly addresses this challenge, offering high cohesive energy that provides inherent reliability against electromigration [5,13], and its chemical stability enables a simplified, barrierless integration scheme [14,15]. The viability of Ru has been well-established, having been previously investigated as a liner material for Cu, where it demonstrated excellent adhesion and gap-filling capabilities [16–19]. Notably, recent studies have confirmed the successful integration of Ru into fabrication processes as a primary conductor [20–22].

While elemental Ru offers a clear path forward, even certain properties (e.g. reported suboptimal adhesion to some dielectrics [15,23]) might hinder its further optimization or tuning for the most scaled and integrated device architecture. The vast design space of binary, ternary, and quaternary compounds thus presents an even greater opportunity to engineer materials with superior properties. By combining elements, it becomes possible to tailor properties beyond what elemental Ru alone can offer, allowing for specific enhancements in electronic transport characteristics, improved interfacial compatibility with diverse dielectric layers, or even stronger interatomic bonding for superior reliability against electromigration.

To identify candidates that not only outperform elemental Cu but also possess robust integration potential, this work extends the conventional high-throughput screening of binary systems to comprehensively explore the expanded material space of ternary and quaternary Ru-based compounds. We employ two key figures of merit: (1) the product of bulk resistivity and mean free path ($\rho_0 \times \lambda$), an intrinsic descriptor of scaling behavior where a lower value predicts more favorable conductivity at reduced dimensions [5,24]; (2) cohesive energy ($E_{\mathrm{coh}}$), which serves as a proxy for reliability, signifying stronger inter-atomic bonding and thus enhanced resistance against electromigration and diffusion [5,25]. Especially, unlike previous studies that often present $\rho_0 \times \lambda$ values by individual categorized each transport direction or focusing solely on the most conductive crystallographic orientation [10,26,27], our work presents an averaged values that more realistically reflects the effective resistance in real interconnects, thereby considering the practical growth environment. This study thereby aims to identify novel compounds predicted to overcome current scaling limitations for future interconnects.

[a] Department of Electronics Engineering, Incheon National University, Yeonsu-gu, Incheon 22012, Republic of Korea.
[b] Major in Advanced Materials and Semiconductor Engineering, Hanyang University, Ansan-si, Gyeonggi-do 15588, South Korea.
[c] Research Institute for Engineering and Technology, Incheon National University, Yeonsu-gu, Incheon 22012, Republic of Korea.

## 2 High-throughput screening

The theoretical framework for this study relies on high-throughput first-principles calculations based on density functional theory (DFT), executed with the Vienna Ab initio Simulation Package (VASP) [28,29]. For these calculations, the exchange-correlation functional was treated with the Perdew-Burke-Ernzerhof (PBE) formulation under the generalized gradient approximation (GGA) [30,31], and ion-electron interactions were described using pseudopotentials generated by the projector-augmented wave (PAW) method [32,33]. The effect of spin polarization was considered in all calculations.

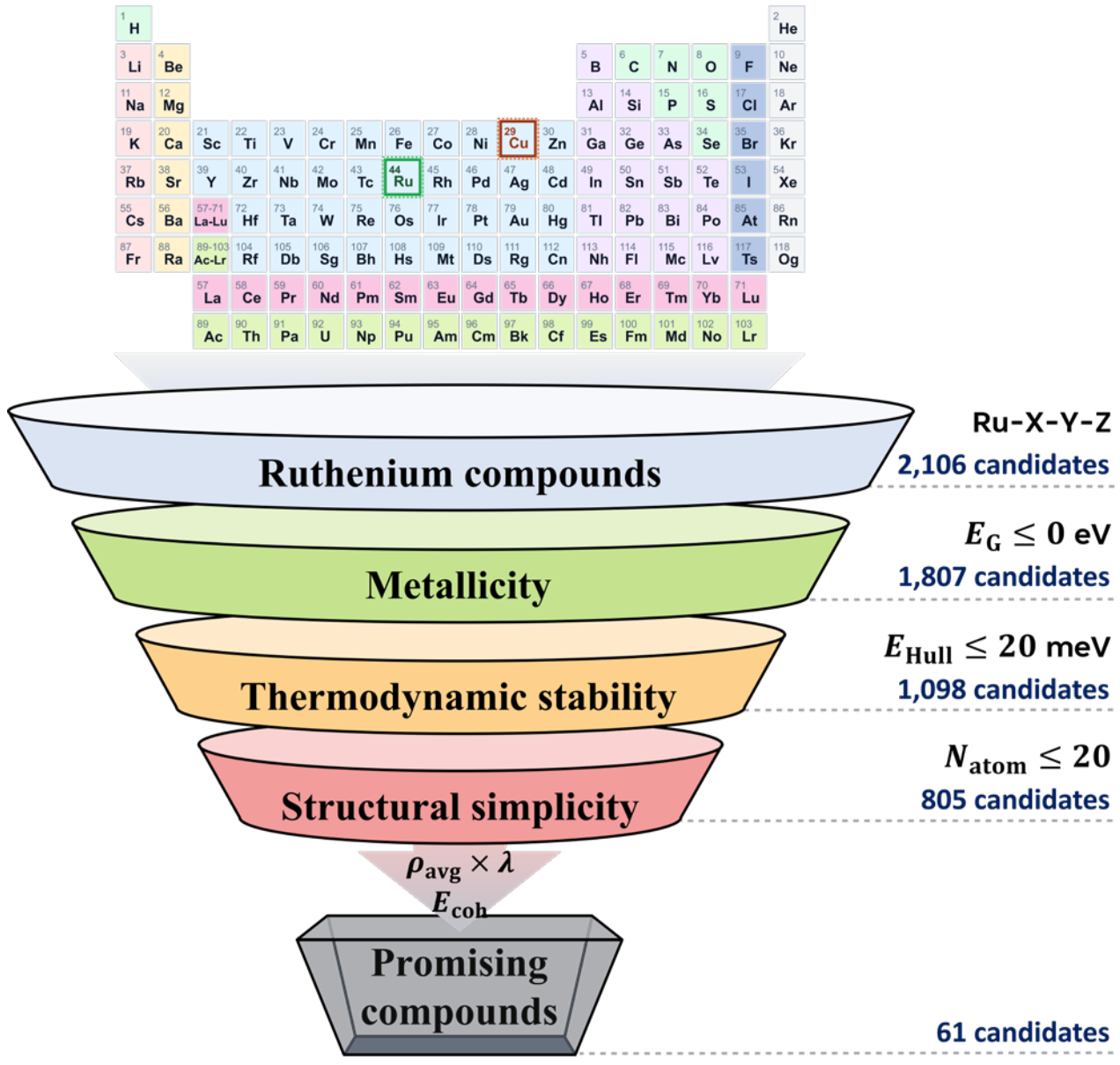

Fig. 1 High-throughput screening procedure for promising ruthenium compounds. Highlighted elements in the periodic table represent copper, the standard interconnect material, and ruthenium, the target element in this study.

To facilitate a broad and efficient material search, we began by sourcing initial crystal structures from the Materials Project database via its Application Programming Interface (API) [34,35]. These candidates were subsequently subjected to a multi-step screening process, depicted in Fig. 1, designed to identify the most promising materials. The screening criteria required that compounds be (1) metallic (band gap $E_G \leq 0$ eV), (2) thermodynamically stable (energy above the convex hull $E_{Hull} \leq 20$ meV/atom), and (3) structurally simple ($N_{atom} \leq 20$ atoms in the primitive unit cell) [36]. Our choice of $E_{Hull} \leq 20$ meV/atom for thermodynamic stability is a deliberate and pragmatically motivated criterion. While strict thermodynamic stability ($E_{Hull} = 0$ eV/atom) is often theoretically preferred, it is known that not all stable compounds are synthesizable, nor all unstable compounds impossible to synthesize [37]. Thus, we selected 20 meV/atom as a threshold slightly above the DFT-calculated known median of metastability of $15 \pm 0.5$ meV/atom [36].

For the selected candidates, static self-consistent field calculations were set up using standardized input parameters generated via the Pymatgen library [38]. To simplify the calculation methodology, we leveraged pre-relaxed structural data directly from the Materials Project database, where geometries are optimized with their convergence criteria (ionic step energy convergence $\leq 5 \times 10^{-4} \times N_{\text{atoms}}$ eV, where $N_{\text{atoms}}$ refers to the number of atoms). Consequently, these optimized structures were used directly for electronic structure calculations without further relaxation. The plane-wave energy cutoff was set to 520 eV with a k-point density of 100/Å$^{-3}$. To properly account for magnetic effects, we directly retrieved and utilized the optimized magnetic ground states for each structure from the Materials Project database. Hubbard $U$ corrections were applied to compounds containing oxygen, following the Materials Project compatibility standards.

Electronic transport properties were evaluated using the semiclassical Boltzmann transport theory implemented in the BoltzTraP2 code [39,40]. To ensure numerical accuracy in the transport integrals, non-self-consistent field calculations were performed on a significantly dense k-point mesh, utilizing at least 50,000 grid points per reciprocal atom. The electrical conductivity tensor $\sigma_{\alpha\beta}$ is defined as:

$$\sigma_{\alpha\beta} = \frac{e^2}{8\pi^3} \sum_n \int_{BZ} d^3k \tau_n(\mathbf{k}) v_{\alpha,n}(\mathbf{k}) v_{\beta,n}(\mathbf{k}) \frac{df_{FD}(E)}{dE} |_{E=E_n(\mathbf{k})} \quad (1)$$

where $\alpha$ and $\beta$ are tensor indices, $e$ is the elementary charge, and $n$ denotes the band index. The integration is performed over the Brillouin zone (BZ), with $\tau_n(\mathbf{k})$ and $v_{\alpha,n}(\mathbf{k})$ representing the carrier relaxation time and the group velocity for each tensor direction, respectively, where $\mathbf{k}$ is the wavevector. The term $df_{FD}(E)/dE$ corresponds to the Fermi window function derived from the Fermi-Dirac distribution $f_{FD}(E)$, and $E_n(\mathbf{k})$ is the eigenenergy. We adopted the constant mean-free-path approximation, where the relaxation time is defined as $\tau_n(\mathbf{k}) = \lambda/|v_n(\mathbf{k})|$. It should be noted that this approximation may not perfectly capture the scattering anisotropies with complex Fermi surface topologies. Nevertheless, previous studies demonstrate that such deviations do not lead to significant errors in the preliminary screening of interconnect candidates [7,41]. Under this assumption, Eq (1) can be reformulated to express conductivity per unit MFP:

$$\frac{\sigma_{\alpha\beta}}{\lambda} = \frac{1}{\rho_{\alpha\beta} \times \lambda} = \frac{e^2}{8\pi^3} \sum_n \int_{BZ} d^3k \frac{v_{\alpha,n}(\mathbf{k}) v_{\beta,n}(\mathbf{k})}{|\mathbf{v}_n(\mathbf{k})|} \frac{df_{FD}(E)}{dE} |_{E=E_n(\mathbf{k})} \quad (2)$$

This derivation indicates that the calculated $\sigma_{\alpha\beta}/\lambda$ is the inverse of the $\rho_{\alpha\beta} \times \lambda$. Therefore, we employ $\rho_{\alpha\beta} \times \lambda$ as the primary intrinsic figure of merit to assess the intrinsic scaling behavior of resistivity, which is particularly relevant for ultranarrow interconnect applications.

## 3. Simulation results and discussion

To evaluate the intrinsic transport properties of the screened compounds, a robust scalar descriptor is required to condense the tensorial outputs of electronic transport calculations. For compounds with cubic symmetry, the isotropic diagonal components allow for a direct scalar representation. However, for non-cubic structures, directional anisotropy necessitates a systematic averaging procedure to ensure a consistent comparison across diverse crystal symmetries. In this work, we primarily define an effective $\rho_{\text{avg}} \times \lambda$ based on a conductivity-averaging scheme. Specifically, a diagonal-only arithmetic average of the three conductivity-proportional components ($\sigma_{xx}/\lambda, \sigma_{yy}/\lambda$, and $\sigma_{zz}/\lambda$) in Eq. (2) is evaluated. This approach represents an effective-medium transport scenario in which multiple crystallographic conduction channels contribute concurrently to the overall electrical response, analogous to a parallel combination of transport pathways. Such a description is appropriate for bulk and polycrystalline materials, where electrical current responds collectively to an applied field rather than being confined to a single crystallographic direction. As a result, the conductivity-based average serves as an effective descriptor of the overall transport performance while minimizing the impact of unfavorable high-resistivity directions [42]. Accordingly, this conductivity-based averaging scheme is adopted throughout the main text to ensure a consistent comparison of materials with varying crystal symmetries in the context of high-throughput screening.

For comparison, an alternative resistivity-based averaging scheme ($\rho^*_{\text{avg}} \times \lambda$) is also considered. In this case, an arithmetic average of the resistivity-proportional components ($\rho_{xx} \times \lambda, \rho_{yy} \times \lambda$, and $\rho_{zz} \times \lambda$) is taken. This averaging method emphasizes the contribution of individual transport directions equally and is therefore more closely associated with transport scenarios in which directional resistances are effectively combined in series [42]. It can be assumed that under the confinement conditions, in narrow wire-like structures, the current is constrained to traverse less favorable crystallographic orientations without the possibility of bypassing high-resistivity pathways. Consequently, this definition highlights the influence of high-resistivity directions and provides an upper bound on the electrical resistivity that a material system may theoretically attain. While this resistivity-based average is valuable for assessing the theoretical limit of transport degradation and for examining directional sensitivity, the conductivity-based averaging scheme is employed as the primary descriptor in this work, as it not only more commonly employed in first-principles transport studies for comparative materials screening [43,7,44] but also more directly comparable to experimentally measured thin-film resistivities. Plots based on the resistivity-based averaging scheme are therefore provided as complementary information in Supplementary Information Section 1 (Figs. S1-S3), while the corresponding values for promising compounds are summarized in the last column of tables.

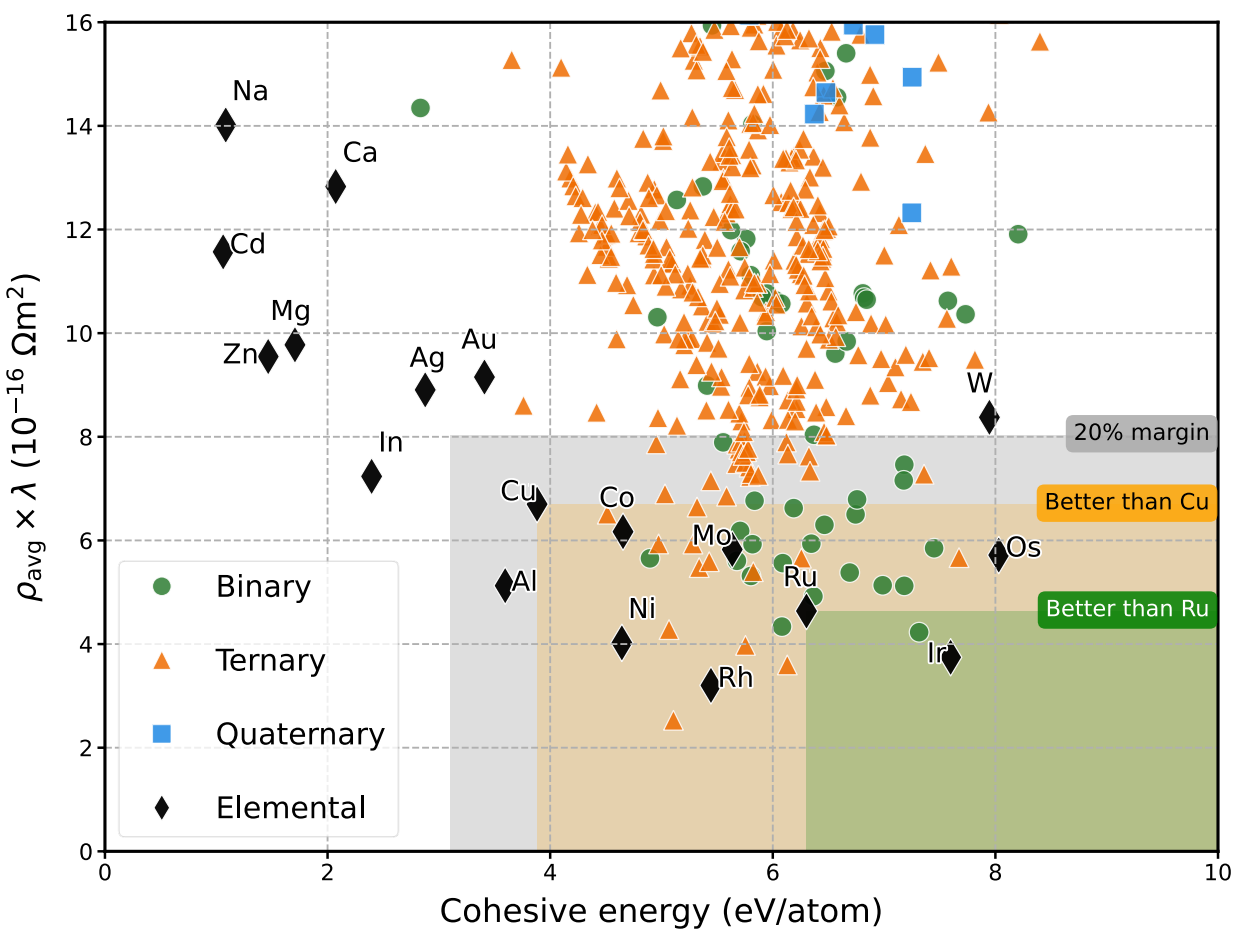

Fig. 2 The electrical properties as a function of cohesive energy for all screened Ru-based compounds. The elemental materials are marked with black diamonds. The shaded areas represent the parameter space where the performance is expected to be superior to each reference material, as well as the performance margin relative to Cu material. Note that compounds with $\rho_{\text{avg}} \times \lambda$ over $16 \times 10^{-16}$ $\Omega\text{m}^2$ are omitted due to their predicted poor electrical performance.

Based on the screening workflow described in Section II, we first examined a total of 2,106 Ru-based compounds, spanning binary, ternary, and quaternary chemistries. By applying a series of screening criteria, the initial materials space was narrowed to 805 metallic and thermodynamically viable phases. These compounds were subsequently evaluated using two key figures of merit: $\rho_{\text{avg}} \times \lambda$ and the cohesive energy $E_{\text{coh}}$. To assess transport performance in a manner relevant to scaled interconnects, elemental Cu was considered as the primary benchmark material. The calculated values for Cu are $\rho_{\text{avg}} \times \lambda = 6.70 \times 10^{-16}\ \Omega\text{m}^2$ and $E_{\text{coh}} = 3.88\ \text{eV}$. Recognizing that actual interconnect properties are governed not only by intrinsic transport metrics but also by, harder-to-quantify factors such as process compatibility, thermal stability, and interfacial adhesion [25,45], we note that materials slightly underperforming Cu in one metric may still offer superior overall integration potential. Furthermore, as our study aims to identify a broad and comprehensive list of promising compounds, considering these complex interdependencies, we allowed candidates up to 20% higher $\rho_{\text{avg}} \times \lambda$, and up to 20% lower $E_{\text{coh}}$ compared to Cu for further consideration. Applying these criteria, a total of 61 promising Ru-based compounds are presented in Fig. 2 Among them, 23 candidates are binary phases, 38 are ternary phases, and no quaternary compounds satisfy the combined transport and stability requirements.

Table 1 Most promising Ru binary compounds

| Chemical formula | Material ID | Crystal symmetry | Experimentally Synthesized [a] | Total magnetization ($\mu_B$) | Cohesive energy [b] (eV) | $\rho_{\alpha\beta} \times \lambda$ ($10^{-16}\ \Omega\,\mathrm{m}^2$) | | | | |
|---|---|---|---|---|---|---|---|---|---|---|
| | | | | | | $\rho_{xx} \times \lambda$ | $\rho_{yy} \times \lambda$ | $\rho_{zz} \times \lambda$ | $\rho_{avg} \times \lambda$ | $\rho^*_{avg} \times \lambda$ |
| AlRu | mp-542569 | Cubic | O [46, 47] | 0 | 5.68 | 5.61 | 5.61 | 5.61 | 5.61 | 5.61 |
| GaRu | mp-22320 | Cubic | O [48] | 0 | 4.90 | 5.65 | 5.65 | 5.65 | 5.65 | 5.65 |
| IrRu | mp-974421 | Hexagonal | | 0 | 6.99 | 5.02 | 5.03 | 5.37 | 5.14 | 5.14 |
| $Ir_2Ru_6$ | mp-862620 | Hexagonal | | 0 | 6.69 | 5.84 | 5.82 | 4.66 | 5.38 | 5.44 |
| $Ir_3Ru$ | mp-974358 | Tetragonal | | 0 | 7.31 | 4.15 | 4.15 | 4.39 | 4.23 | 4.23 |
| LuRu | mp-11495 | Cubic | O [49] | 0 | 5.84 | 6.77 | 6.77 | 6.77 | 6.77 | 6.77 |
| $Mn_2Ru_6$ | mp-865045 | Hexagonal | | 1.179 | 5.55 | 8.08 | 8.08 | 7.55 | 7.89 | 7.90 |
| $Mo_2Ru_6$ | mp-975834 | Hexagonal | | 0 | 6.19 | 9.21 | 9.21 | 4.24 | 6.62 | 7.55 |
| OsRu | mp-1220023 | Hexagonal | | 0 | 7.18 | 5.62 | 5.62 | 4.35 | 5.12 | 5.20 |
| $Os_2Ru_6$ | mp-974326 | Hexagonal | | 0 | 6.74 | 5.92 | 5.93 | 8.08 | 6.50 | 6.64 |
| $Re_3Ru$ | mp-974455 | Orthorhombic | | 0 | 7.18 | 6.08 | 7.41 | 9.76 | 7.46 | 7.75 |
| $Re_6Ru_2$ | mp-974625 | Hexagonal | | 0 | 7.18 | 8.52 | 8.53 | 5.42 | 7.16 | 7.49 |
| $RuRh_4$ | mp-1219522 | Trigonal | | 0.001 | 6.08 | 6.31 | 3.75 | 3.75 | 4.34 | 4.61 |
| $Ru_6Rh_2$ | mp-1186926 | Hexagonal | | 0 | 6.09 | 5.95 | 5.95 | 4.92 | 5.56 | 5.61 |
| $Ru_6W_2$ | mp-862655 | Hexagonal | | 0 | 6.76 | 9.41 | 9.41 | 4.37 | 6.79 | 7.73 |
| ScRu | mp-30867 | Cubic | O [50, 51] | 0 | 5.71 | 6.18 | 6.18 | 6.18 | 6.18 | 6.18 |
| TaRu | mp-1601 | Tetragonal | O [52, 53] | 0 | 7.45 | 6.48 | 6.48 | 4.89 | 5.85 | 5.95 |
| TcRu | mp-1217363 | Hexagonal | | 0 | 6.37 | 5.25 | 5.25 | 4.37 | 4.92 | 4.96 |
| $Tc_2Ru_6$ | mp-867356 | Hexagonal | | 0 | 6.37 | 8.51 | 8.51 | 7.24 | 8.04 | 8.09 |
| $Tc_6Ru_2$ | mp-861630 | Hexagonal | | 0 | 6.35 | 7.08 | 7.09 | 4.49 | 5.94 | 6.22 |
| $URu_3$ | mp-1263 | Cubic | O [54, 55] | 0.003 | 6.46 | 6.30 | 6.30 | 6.30 | 6.30 | 6.30 |
| VRu | mp-1395 | Cubic | O [56, 57] | 0.559 | 5.80 | 5.31 | 5.31 | 5.33 | 5.32 | 5.32 |
| YbRu | mp-567116 | Cubic | O [58] | 0 | 5.82 | 5.93 | 5.93 | 5.93 | 5.93 | 5.93 |

[a] The synthesizability of the materials was determined by referencing the compositions provided in the ICSD.

[b] The cohesive energies are obtained from the Materials Project database.

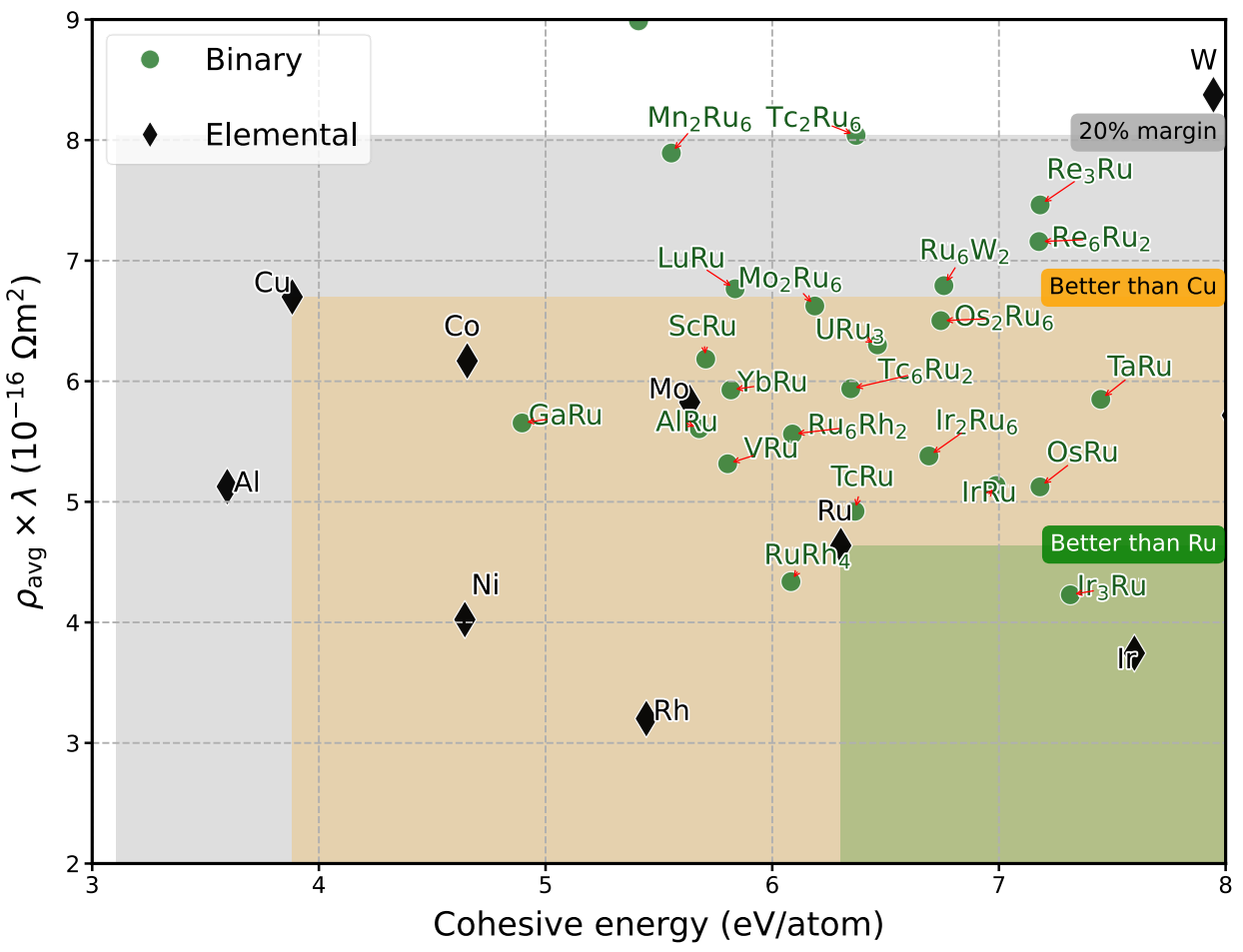

Fig. 3 Potential candidates among binary Ru compounds against Cu. The highlighted zones (orange and green) categorize compounds exceeding the performance of elemental Cu and Ru, respectively. A total of 23 binary candidates is identified within the 20% margin of the Cu, representing the most promising system for advanced interconnect material. Elemental benchmarks (Cu, Ru, etc.) are provided for comparative reference.

The comprehensive screening results for binary Ru-based compounds are summarized in Fig. 3 The shaded regions delineate performance benchmarks defined by elemental Cu, the current industry standard, and elemental Ru, which serves as a reference material in this study. As discussed above, the grey-shaded region represents an extended benchmark window introduced to avoid over-restrictive filtering based solely on intrinsic transport and cohesive energy metrics. This relaxed window is intended to preserve candidates that may exhibit advantageous secondary properties not explicitly captured by the present descriptors. Out of 93 binary compounds that satisfied our screening workflow, 23 compounds were identified within the defined performance window. Among these binary candidates, eight compounds—AlRu [46,47], GaRu [48], LuRu [49], ScRu [50,51], TaRu [52,53], $URu_3$ [54,55], VRu [56,57], and YbRu [58]—have been reported to be experimentally synthesizable, as verified through experimental reports within the Inorganic Crystal Structure Database (ICSD). Notably, AlRu has also attracted recent attention as a potential alternative material for interconnect technology, having been experimentally evaluated as promising in prior research [59]. Detailed calculated properties and structural information for all selected binary compounds are provided in Table 1.

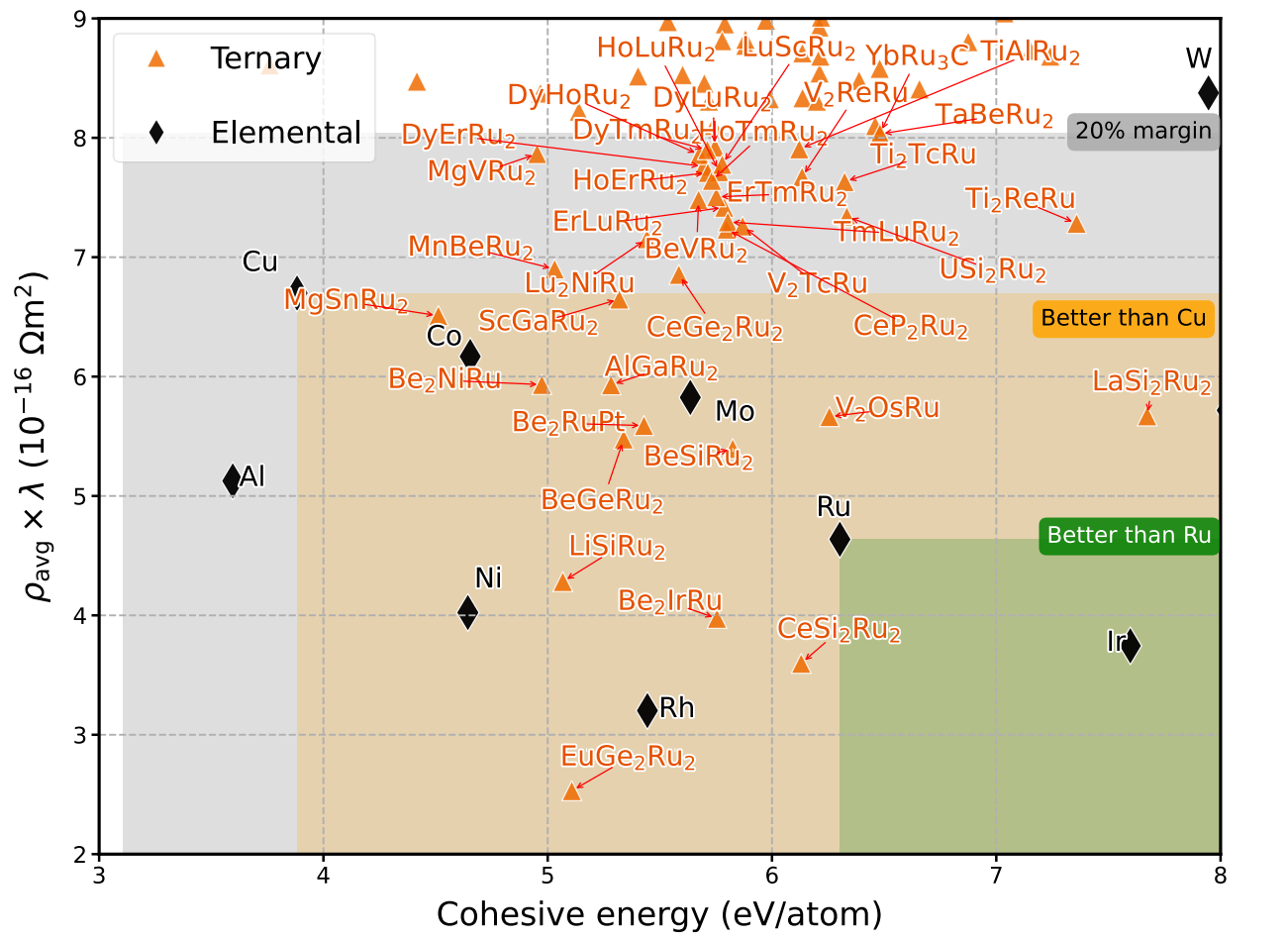

Fig. 4 High-throughput screening of ternary Ru compounds. The scatter plot maps the distribution of ternary candidates across the performance-stability design space. The shaded regions indicate the performance thresholds relative to each elemental property.

Fig. 4 presents the corresponding screening results for ternary Ru-based compounds. From an initial pool of 626 ternary phases, 38 compounds were identified within the defined performance window. Among the selected ternary candidates, six compounds— $CeGe_2Ru_2$ [60,61], $CeP_2Ru_2$ [62], $CeSi_2Ru_2$ [63,64], $EuGe_2Ru_2$ [65,66], $LaSi_2Ru_2$ [63], and $USi_2Ru_2$ [67,68]—have been experimentally synthesized. Attempts to utilize these materials for interconnect applications have not yet been documented in the literature. The full list of ternary candidates and their calculated properties are summarized in Table 2. In contrast to the binary and ternary systems, none of the quaternary Ru-based compounds fall within our performance window. This absence is largely a consequence of the screening criteria applied during the initial screening stages rather than an inherent lack of electrical performance in complex compounds. As the number of constituent elements increases from binary to quaternary systems, the number of atoms constituting the primitive unit cell tends to increase to accommodate the required stoichiometry. Statistically, while only ~10% of binary compounds exceed the $N_{\text{atom}}$ limit, this fraction rises to ~25% for ternary systems and drastically increases to ~51% for quaternary systems. Consequently, a significant portion of quaternary candidates were excluded prior to the transport evaluation due to this constraint.

Despite the identification of several promising Ru-based compounds within the defined performance window, safety and material feasibility considerations must be addressed for specific candidates. In particular, compounds containing beryllium (Be) require careful handling due to its well-established carcinogenicity, while technetium (Tc) and uranium (U) containing materials raise concerns associated with radioactivity. While these factors certainly pose significant challenges for conventional semiconductor manufacturing, their inclusion in this foundational screening is justified by the potential for unique material properties that may warrant their use in specialized applications and exhibit irreplaceable performance. Indeed, despite its carcinogenicity, Be is utilized in various field such as aviation, aerospace, and nuclear fusion due to its excellent properties. Similarly, radioactive materials containing Tc and U can also still be considered for use under strict safety measures when their superior properties justify their application in specialized fields. However, the use of such materials necessitates high-level engineering controls, comprehensive personal protective equipment, continuous environmental monitoring, and through training on save handling procedures [69–72]. For the primary objective of this study, which is to provide a broad and exhaustive list of potentially useful materials for subsequent research, these data were included without arbitrary exclusion. Furthermore, while the number of compounds with confirmed experimental synthesizability may not be extensive, their selection based on a thermodynamic stability criterion ($E_{\text{hull}} \leq 20$ meV/atom) suggests a high potential for future experimental synthesis.

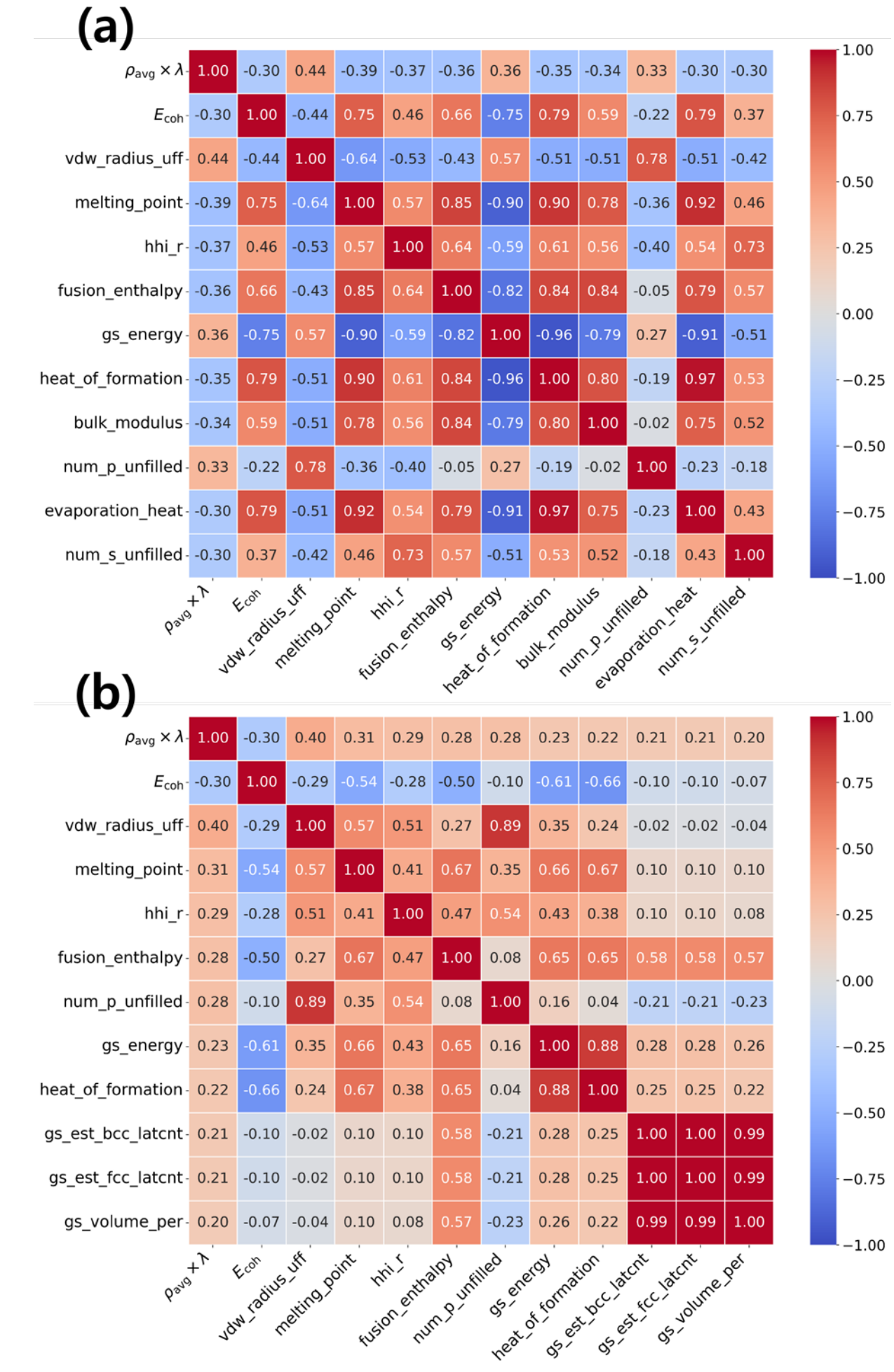

Fig. 5 Pearson correlation heatmap of compositional descriptors for binary compounds. (a) the matrix illustrates the correlation coefficients between the top 10 average-related compositional features extracted via the XenonPy toolkit and the primary screening metrics: $\rho_0 \times \lambda$ and $E_{\text{coh}}$. (b) The matrix illustrates the correlation coefficients between the top 10 variance-related compositional features and the same screening metrics.

Table 2 Most promising Ru ternary compounds

| Chemical formula | Material ID | Crystal symmetry | Experimentally Synthesized [a] | Total magnetization ($\mu_B$) | Cohesive energy [b] (eV) | $\rho_{\alpha\beta} \times \lambda$ $(10^{-16}\ \Omega\ m^2)$ | | | | |
|---|---|---|---|---|---|---|---|---|---|---|
| | | | | | | $\rho_{xx} \times \lambda$ | $\rho_{yy} \times \lambda$ | $\rho_{zz} \times \lambda$ | $\rho_{avg} \times \lambda$ | $\rho^*_{avg} \times \lambda$ |
| $AlGaRu_2$ | mp-1183218 | Cubic | | 0 | 5.28 | 5.93 | 5.93 | 5.93 | 5.93 | 5.93 |
| $BeGeRu_2$ | mp-862713 | Cubic | | 0 | 5.34 | 5.48 | 5.48 | 5.48 | 5.48 | 5.48 |
| $BeSiRu_2$ | mp-867835 | Cubic | | 0 | 5.83 | 5.40 | 5.40 | 5.40 | 5.40 | 5.40 |
| $BeVRu_2$ | mp-867874 | Cubic | | 0 | 5.67 | 7.48 | 7.48 | 7.48 | 7.48 | 7.48 |
| $Be_2IrRu$ | mp-1183419 | Cubic | | 0 | 5.75 | 3.97 | 3.97 | 3.97 | 3.97 | 3.97 |
| $Be_2NiRu$ | mp-1183421 | Cubic | | 0 | 4.97 | 5.93 | 5.93 | 5.93 | 5.93 | 5.93 |
| $Be_2RuPt$ | mp-865021 | Cubic | | 0 | 5.43 | 5.59 | 5.59 | 5.59 | 5.59 | 5.59 |
| $CeGe_2Ru_2$ | mp-22343 | Tetragonal | O [60, 61] | 0.594 | 5.58 | 9.36 | 8.77 | 4.61 | 6.85 | 7.58 |
| $CeP_2Ru_2$ | mp-574244 | Tetragonal | O [62] | 0.428 | 5.80 | 9.32 | 8.96 | 5.11 | 7.23 | 7.80 |
| $CeSi_2Ru_2$ | mp-3566 | Tetragonal | O [63, 64] | 0.242 | 6.13 | 4.83 | 4.54 | 2.46 | 3.60 | 3.94 |
| $DyErRu_2$ | mp-1183783 | Cubic | | 0 | 5.69 | 7.76 | 7.76 | 7.76 | 7.76 | 7.76 |
| $DyHoRu_2$ | mp-1183786 | Cubic | | 0 | 5.68 | 7.86 | 7.86 | 7.86 | 7.86 | 7.86 |
| $DyLuRu_2$ | mp-1183818 | Cubic | | 0 | 5.75 | 7.92 | 7.92 | 7.92 | 7.92 | 7.92 |
| $DyTmRu_2$ | mp-1184074 | Cubic | | 0 | 5.71 | 7.90 | 7.90 | 7.90 | 7.90 | 7.90 |
| $ErLuRu_2$ | mp-1184363 | Cubic | | 0 | 5.79 | 7.42 | 7.42 | 7.42 | 7.42 | 7.42 |
| $ErTmRu_2$ | mp-1184242 | Cubic | | 0 | 5.75 | 7.51 | 7.51 | 7.51 | 7.51 | 7.51 |
| $EuGe_2Ru_2$ | mp-21417 | Tetragonal | O [65, 66] | 6.389 | 5.11 | 3.64 | 3.37 | 1.63 | 2.53 | 2.88 |
| $HoErRu_2$ | mp-976311 | Cubic | | 0 | 5.71 | 7.71 | 7.70 | 7.72 | 7.71 | 7.71 |
| $HoLuRu_2$ | mp-973114 | Cubic | | 0 | 5.77 | 7.71 | 7.71 | 7.72 | 7.71 | 7.71 |
| $HoTmRu_2$ | mp-1184831 | Cubic | | 0 | 5.73 | 7.64 | 7.64 | 7.64 | 7.64 | 7.64 |
| $LaSi_2Ru_2$ | mp-5105 | Tetragonal | O [63] | 0 | 7.67 | 7.57 | 7.14 | 3.89 | 5.67 | 6.20 |
| $LiSiRu_2$ | mp-865838 | Cubic | | 0.001 | 5.07 | 4.28 | 4.28 | 4.28 | 4.28 | 4.28 |
| $LuScRu_2$ | mp-973433 | Cubic | | 0 | 5.78 | 7.78 | 7.78 | 7.78 | 7.78 | 7.78 |
| $Lu_2NiRu$ | mp-865335 | Cubic | | 0 | 5.44 | 7.02 | 7.22 | 7.22 | 7.15 | 7.15 |
| $MgSnRu_2$ | mp-1185980 | Cubic | | 0 | 4.51 | 6.51 | 6.51 | 6.51 | 6.51 | 6.51 |
| $MgVRu_2$ | mp-1185620 | Cubic | | 0.001 | 4.95 | 7.87 | 7.87 | 7.87 | 7.87 | 7.87 |
| $MnBeRu_2$ | mp-1185984 | Cubic | | 1.989 | 5.03 | 6.90 | 6.90 | 6.90 | 6.90 | 6.90 |
| $ScGaRu_2$ | mp-867156 | Cubic | | 0.004 | 5.32 | 6.65 | 6.65 | 6.65 | 6.65 | 6.65 |
| $TaBeRu_2$ | mp-867114 | Cubic | | 0 | 6.48 | 8.03 | 8.03 | 8.03 | 8.03 | 8.03 |
| $TiAlRu_2$ | mp-866155 | Cubic | | 0 | 6.12 | 7.91 | 7.91 | 7.91 | 7.91 | 7.91 |
| $Ti_2ReRu$ | mp-972275 | Cubic | | 0 | 7.36 | 7.28 | 7.28 | 7.28 | 7.28 | 7.28 |
| $Ti_2TcRu$ | mp-865650 | Cubic | | 0 | 6.32 | 7.63 | 7.63 | 7.63 | 7.63 | 7.63 |
| $TmLuRu_2$ | mp-983312 | Cubic | | 0 | 5.80 | 7.29 | 7.30 | 7.30 | 7.30 | 7.30 |
| $USi_2Ru_2$ | mp-3388 | Tetragonal | O [67, 68] | 1.639 | 6.33 | 9.28 | 8.82 | 5.33 | 7.34 | 7.81 |
| $V_2OsRu$ | mp-971737 | Cubic | | 0.249 | 6.26 | 5.66 | 5.66 | 5.66 | 5.66 | 5.66 |
| $V_2ReRu$ | mp-981365 | Cubic | | 0.003 | 6.13 | 7.67 | 7.67 | 7.67 | 7.67 | 7.67 |
| $V_2TcRu$ | mp-865501 | Cubic | | 0.007 | 5.87 | 7.26 | 7.26 | 7.26 | 7.26 | 7.26 |
| $YbRu_3C$ | mp-1206413 | Cubic | | 0 | 6.48 | 8.04 | 8.04 | 8.04 | 8.04 | 8.04 |

[a] The synthesizability of the materials was determined by referencing the compositions provided in the ICSD.

[b] The cohesive energies are obtained from the Materials Project database.

From a performance standpoint, our simulations indicate that, among the screened binary systems, only $Ir_3Ru$ marginally outperforms elemental Ru, while all other candidates fall short of surpassing bulk Ru in the metrics considered here. A similar trend is observed for ternary systems, where none achieve performance exceeding that of elemental Ru. This observation is consistent with prior studies on Cobalt (Co) based binary interconnect materials, which likewise reported few compounds ($Co_3Ni$, CoPt, and $CoPt_3$) outperforming elemental Co when evaluated using comparable intrinsic transport related criteria [73]. This overall trend suggests that, when assessed solely through intrinsic electronic transport indicators, it is intrinsically challenging for compound materials to simultaneously exceed the performance of their best elemental counterparts. The observed slight advantage of $Ir_3Ru$ over elemental Ru, for instance, is highly influenced by the inherently superior electronic transport properties of its constituent Ir. Indeed, a broader analysis encompassing all metallic elemental materials alongside our binary compounds (Supplementary Fig. S4) reveals that compound transport properties are often positioned between those of their constituent elements. While some compounds like $Ru_6Rh_2$, $Tc_6Ru_2$, $Re_3Ru$, and AlRu exhibit higher resistivity factor than either constituent element, others such as $Ru_6W_2$, $RuRh_4$, and TcRu demonstrate properties falling between the electrical characteristics of their respective elemental components. This spectrum of performance analytically demonstrates that while complex atomic and electronic interactions are at play, compound formation predominantly redistributes, rather than inherently transcends, the intrinsic transport properties relative to their elemental building blocks.

To elucidate the fundamental physical origins of this systematic performance trend, we focused on binary compounds as an ideal model system. This targeted approach allows us to precisely isolate the intrinsic structure-property relationships governing A-B interactions, avoiding the confounding effects of multi-element structural complexity inherent in ternary and quaternary systems. Using the XenonPy toolkit[74], we extracted 116 compositional descriptors, focusing on both the weighted average and weighted variance features to comprehensively capture the effective system properties and the degree of elemental dissimilarity within each compound. As shown in the Pearson correlation heatmaps (Fig. 5a), the average-related descriptors reveal that the effective van der Waals radius based on the universal force field (vdw_radius_uff) exhibits the positive correlation with $\rho_{avg} \times \lambda$ ($r = 0.44$) (Detailed descriptions of the other variance descriptors are provided in the Supplementary Information Table S1). This indicates that as the overall cell size of the compound increases, the intrinsic transport performance progressively degrades. Furthermore, this data reveals strong positive correlations ($r \geq 0.75$) between $E_{coh}$ and thermodynamic stability metrics such as melting_point, evaporation_heat, and heat_of_formation. Since the melting point is widely used as a proxy for electromigration performance [75,76], this strong correlation indicates that compounds with higher cohesive energy inherently possess enhanced interconnect reliability.

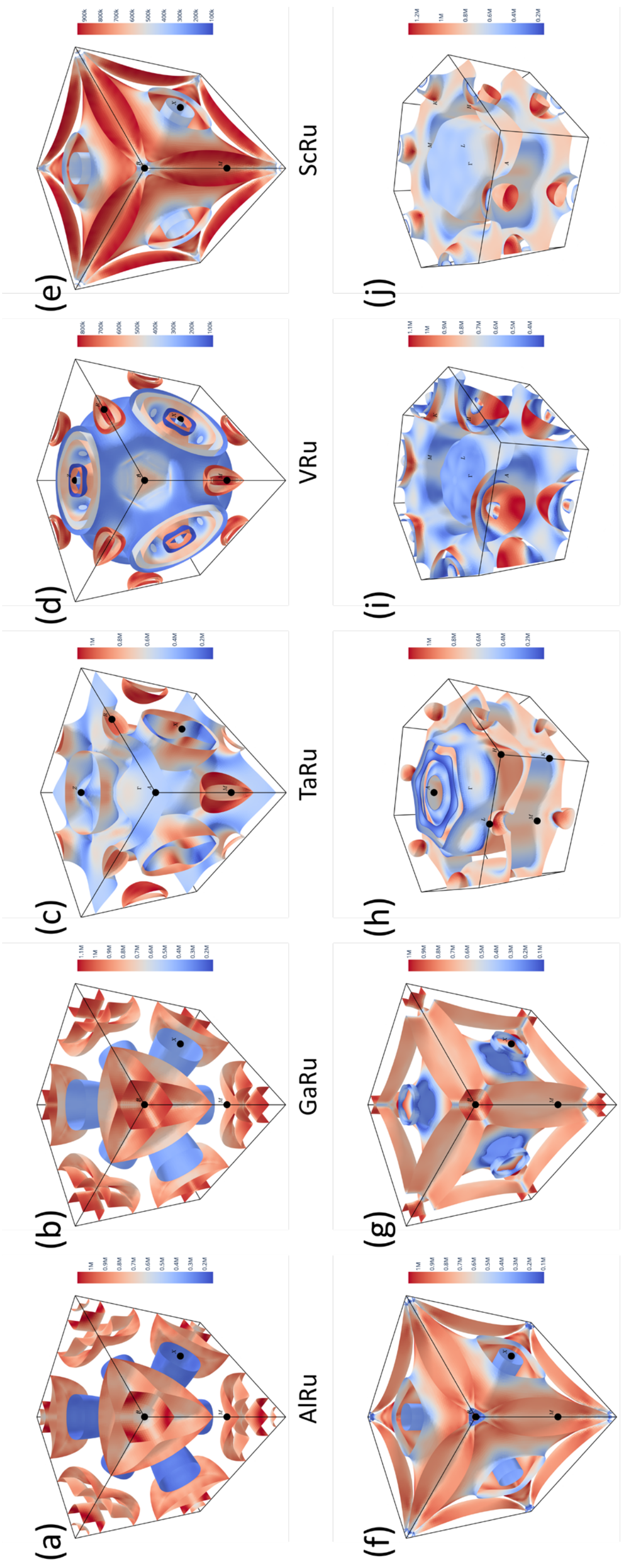

Fig. 6. Fermi surface topologies of selected binary Ru compounds. The calculated Fermi surface is categorized by their underlying crystal symmetry: (a-g) face-centered cubic (FCC) derived structures and (h-j) hexagonal close-packed (HCP) derived structures. The color mapping indicates the Fermi velocity distribution on the surface.

The trend of increasing $\rho_{avg} \times \lambda$ with larger effective atomic radii can be fundamentally rationalized through the relationship between real-space lattice dimensions and reciprocal-space electronic structure. While compound formation inherently modifies the periodic atomic potential and can alter the total valence electron concentration, the geometric volume effect plays a dominant role when comparing materials with similar Fermi surface topologies. Under the condition that the fundamental band structure is largely preserved, the incorporation of larger atoms inherently expands the real-space lattice parameters, which subsequently decreases the volumetric electron density. According to Luttinger's theorem [77,78], the volume enclosed by the Fermi surface is strictly proportional to this particle density. Consequently, this real-space expansion inversely contracts the reciprocal-space Brillouin zone, which inherently bounds and effectively reduces the cross-sectional area of the Fermi surface [79]. Given that the effective Fermi surface area correlates with electrical conductivity [80], this geometric contraction provides a compelling physical rationale for how increased cell volume degrades intrinsic transport performance. In addition to the average structural dimensions, we evaluated the variance descriptors to understand the impact of elemental size mismatch (Fig. 5b). The variance in the vdw_radius_uff shows a positive correlation ($r = 0.40$) with $\rho_{\mathrm{avg}} \times \lambda$. This analytically demonstrates that compounds formed between elements with a large atomic size discrepancy tend to exhibit higher resistivity factors. Indeed, referring back to the distribution of binary compounds in Fig. 3, constituents with atomic radii similar to Ru—predominantly those located nearby in the periodic table such as $\mathrm{Ir}$, $\mathrm{Rh}$, $\mathrm{Os}$ and $\mathrm{Tc}$—exhibited superior transport characteristics, corroborating this descriptor trend.

These theoretical rationales are directly supported by our electronic structure analysis. Notably, compounds sharing the same crystal structure and similar Fermi surface topology exhibit closely grouped $\rho_{\mathrm{avg}} \times \lambda$ values. For instance, as illustrated in Fig. 6, among 1:1 binary compounds, $\mathrm{AlRu}$, $\mathrm{GaRu}$, and $\mathrm{TaRu}$ share a face-centered cubic (FCC)-derived structure with analogous Fermi surface morphologies and accordingly cluster within a comparable $\rho_{\mathrm{avg}} \times \lambda$ range. Similarly, $\mathrm{ScRu}$, $\mathrm{LuRu}$, and $\mathrm{YbRu}$ also adopt an FCC-derived structure with mutually similar, yet distinctly different Fermi surface morphologies from the former group, resulting in their own closely grouped $\rho_{\mathrm{avg}} \times \lambda$ values. However, even within this isostructural group, compounds incorporating larger atoms (e.g., $\mathrm{Lu}$ and $\mathrm{Yb}$) exhibit higher values, which perfectly aligns with our average cell expansion analysis. Conversely, $\mathrm{VRu}$, which adopts the same FCC structure but exhibits a distinctly different Fermi surface morphology, displays a somewhat different transport value. An analogous topological clustering is observed among hexagonal close-packed (HCP) structured compounds: $\mathrm{IrRu}$, $\mathrm{OsRu}$, and $\mathrm{TcRu}$ share highly similar Fermi surface features and correspondingly cluster within a very narrow, highly conductive range. Consequently, from a practical material design perspective, selecting constituent elements with similar atomic radii to minimize lattice mismatch, while simultaneously keeping the overall cell volume small to prevent Brillouin zone contraction, represents a highly effective strategy for minimizing intrinsic transport degradation during compound formation or alloying.

While elemental metals often define upper bounds in intrinsic performance metrics, compound formation can offer distinct integration-driven advantages that are not captured by these indicators alone. In the context of interconnect technology, prior studies have demonstrated that specific compound phases can exhibit superior adhesion to dielectrics, suppress atomic diffusion, and enable self-forming barrier or liner behaviors [81–86]. These attributes are critical for enhancing reliability and allowing for reduced barrier/liner thickness-a key requirement at extremely scaled dimensions where interfacial effects dominate the effective resistance. Moreover, exploring compound compositions substantially expands the accessible materials space beyond the limited set of elemental conductors, enabling the identification of candidates that strike a more favorable balance between intrinsic electrical performance and integration-relevant feasibility. In this context, the present results provide a meaningful framework for guiding future interconnect material exploration beyond elemental systems.

## Conclusions

In this study, we conducted a comprehensive high-throughput screening of Ru-based binary, ternary, and quaternary compounds to identify promising candidates for next-generation interconnect applications. By evaluating 2,106 initial phases based on first-principles calculations, we identified 61 viable candidates—23 binary and 38 ternary compounds—that satisfy rigorous stability and performance criteria relative to the industry-standard Cu. Our analysis highlights several experimentally synthesized candidates, including the binary compound $\mathrm{AlRu}$, which has already been investigated as a potential interconnect material. Regarding quaternary systems, although no candidates were identified within our specific screening window, this outcome is significantly influenced by the rigorous structural constraints applied during our screening process, where a substantial number of candidates were preemptively excluded due to the increased stoichiometric complexity inherent in moving from binary to quaternary systems. From a performance standpoint, our results indicate that while elemental Ru sets a high benchmark in intrinsic electronic transport, compound formation offers a pathway to engineer materials with distinct advantages. With the exception of $\mathrm{Ir_3Ru}$, most screened compounds do not surpass the performance distribution of elemental Ru; however, the value of compound formation extends beyond simple resistivity minimization. The screened compounds offer potential integration-driven advantages, such as enhanced adhesion, chemical stability, and the possibility of barrierless integration, which are essential for minimizing the effective resistance in sub-nanometer interconnect stacks. Furthermore, our descriptor and Fermi surface topology analyses reveal that the overall cell volume is one of the primary structural factors governing intrinsic transport degradation upon compound formation. Additionally, as observed in compounds formed with elements adjacent to Ru in the periodic table—such as Ir, Os, Rh, and Tc—minimizing the atomic size mismatch between constituent elements can further mitigate transport degradation, providing an effective design principle for screening next-generation interconnect compounds. Consequently, this study provides a foundational roadmap for expanding the materials design space, guiding the industry transition from elemental conductors to optimized compound systems that balance intrinsic performance with process feasibility. Future investigations will focus on providing a more comprehensive understanding of transport properties under realistic device geometries and confinement conditions. Integrating additional scattering mechanisms, such as grain boundary and surface scattering, will be crucial, particularly given their pronounced impact on thin film resistivity. Leveraging advanced computational frameworks like EPW [87] or PERTURBO[88] to quantify these complex scattering phenomena, including electron-phonon interactions, will significantly advance the performance classification and practical application of these materials as next-generation interconnects.

## Author contributions

**Gyungho Maeng:** Data curation, Formal analysis, Investigation, Methodology, Software, Validation, Visualization, Writing – original draft. **Subeen Lim:** Formal analysis, Methodology, **Bonggeun Shong:** Funding acquisition, Validation, Writing – review & editing, and **Yeonghun Lee:** Supervision, Conceptualization, Project administration, Formal analysis, Funding acquisition, Methodology, Validation, Writing – review & editing.

## Conflicts of interest

The authors declare that they have no known competing financial interest or personal relationships that could have appeared to influence the work reported in this paper.

## Data availability

All data supporting the findings of this study have been obtained by the authors and presented within the main text and supplementary information (SI). Additional data are available from the corresponding author upon reasonable request.

## Acknowledgements

This work was supported by the Technology Innovation Program [Public-private joint investment semiconductor R&D program (K-CHIPS) to foster high-quality human resources] [No. RS-2023-00236667, High performance Ru-TiN interconnects via high temperature atomic layer deposition (ALD) and development on new interconnect materials based on ALD] funded by the Ministry of Trade, Industry & Energy (MOTIE, Korea) (No. 1415187401). This work was supported by the National Supercomputing Center with supercomputing resources including technical support (No. KSC-2025-CRE-0054).

## References

1 K. Fuchs, *Math. Proc. Camb. Phil. Soc.*, 1938, **34**, 100–108.
2 E. H. Sondheimer, *Advances in Physics*, 2001, **50**, 499–537.
3 A. F. Mayadas and M. Shatzkes, *Phys. Rev. B*, 1970, **1**, 1382–1389.
4 T. Markussen, S. Aboud, A. Blom, N. A. Lanzillo, T. Gunst, J. Cobb, T. M. Philip and R. R. Robison, in *2020 IEEE International Interconnect Technology Conference (IITC)*, IEEE, San Jose, CA, USA, 2020, pp. 76–78.
5 D. Gall, *Journal of Applied Physics*, 2020, **127**, 050901.
6 H.-D. Liu, Y.-P. Zhao, G. Ramanath, S. P. Murarka and G.-C. Wang, *Thin Solid Films*, 2001, **384**, 151–156.
7 S. Dutta, K. Sankaran, K. Moors, G. Pourtois, S. Van Elshocht, J. Bömmels, W. Vandervorst, Z. Tőkei and C. Adelmann, *Journal of Applied Physics*, 2017, **122**, 025107.
8 S. Dutta, K. Moors, M. Vandemaele and C. Adelmann, *IEEE Electron Device Lett.*, 2018, **39**, 268–271.
9 J. H. Moon, S. Kim, T. Kim, Y. S. Jeon, Y. Kim, J.-P. Ahn and Y. K. Kim, *Journal of Materials Science & Technology*, 2022, **105**, 17–25.
10 D. Gall, *Journal of Applied Physics*, 2016, **119**, 085101.
11 P. Kapur, J. P. McVittie and K. C. Saraswat, *IEEE Trans. Electron Devices*, 2002, **49**, 590–597.
12 R. Saligram, S. Datta and A. Raychowdhury, *IEEE Trans. Circuits Syst. I*, 2022, **69**, 4610–4618.
13 M. H. van der Veen, N. Heyler, O. V. Pedreira, I. Ciofi, S. Decoster, V. V. Gonzalez, N. Jourdan, H. Struyf, K. Croes, C. J. Wilson and Zs. Tőkei, in *2018 IEEE International Interconnect Technology Conference (IITC)*, 2018, pp. 172–174.
14 Liang Gong Wen, C. Adelmann, O. V. Pedreira, S. Dutta, M. Popovici, B. Briggs, N. Heylen, K. Vanstreels, C. J. Wilson, S. Van Elshocht, K. Croes, J. Bommels and Z. Tokei, in *2016 IEEE International Interconnect Technology Conference / Advanced Metallization Conference (IITC/AMC)*, IEEE, San Jose, CA, USA, 2016, pp. 34–36.
15 Y. Kotsugi, S.-M. Han, Y.-H. Kim, T. Cheon, D. K. Nandi, R. Ramesh, N.-K. Yu, K. Son, T. Tsugawa, S. Ohtake, R. Harada, Y.-B. Park, B. Shong and S.-H. Kim, *Chem. Mater.*, 2021, **33**, 5639–5651.
16 C.-C. Yang, T. Spooner, S. Ponoth, K. Chanda, A. Simon, C. Lavoie, M. Lane, C.-K. Hu, E. Liniger, L. Gignac, T. Shaw, S. Cohen, F. McFeely and D. Edelstein, in *2006 International Interconnect Technology Conference*, IEEE, Burlingame, CA, 2006, pp. 187–190.
17 H. Y. Huang, C. H. Hsieh, S. M. Jeng, H. J. Tao, M. Cao and Y. J. Mii, in *2010 IEEE International Interconnect Technology Conference*, IEEE, Burlingame, CA, USA, 2010, pp. 1–3.
18 J. Rullan, T. Ishizaka, F. Cerio, S. Mizuno, Y. Mizusawa, T. Ponnuswamy, J. Reid, A. McKerrow and C.-C. Yang, in *2010 IEEE International Interconnect Technology Conference*, IEEE, Burlingame, CA, USA, 2010, pp. 1–3.
19 C.-C. Yang, S. Cohen, T. Shaw, P.-C. Wang, T. Nogami and D. Edelstein, *IEEE Electron Device Lett.*, 2010, **31**, 722–724.
20 D. Wan, S. Paolillo, N. Rassoul, B. K. Kotowska, V. Blanco, C. Adelmann, F. Lazzarino, M. Ercken, G. Murdoch, J. Bömmels, C. J. Wilson and Z. Tökei, in *2018 IEEE International Interconnect Technology Conference (IITC)*, 2018, pp. 10–12.
21 C. Penny, K. Motoyama, S. Ghosh, T. Bae, N. Lanzillo, S. Sieg, C. Park, L. Zou, H. Lee, D. Metzler, J. Lee, S. Cho, M. Shoudy, S. Nguyen, A. Simon, K. Park, L. Clevenger, B. Anderson, C. Child, T. Yamashita, J. Arnold, T. Wu, T. Spooner, K. Choi, K.-I. Seo and D.

Guo, in *2022 International Electron Devices Meeting (IEDM)*, IEEE, San Francisco, CA, USA, 2022, p. 12.1.1-12.1.4.
22 A. Dutta, A. Peer, C. Jezewski, S. Siddiqui, I. Jenkins, E. Khora, G. Auluck, Y. Huang, F. Bedoya, N. Kabir, S. Mocherla, P. R. Saha, L. Shoer, K. Chan, A. Tanneeru, J. Gupta, V. B. Jeevendrakumar, D. Collins, S. Madhusoodhanan, J. Bielefeld, W. Brezinski, R. Fayad, S. Sukrittanon, S. Naskar, S. Kosaraju, N. Nair, G. Singh, J. D. Silva, C. Engel, N. Franco, B. Krist, J. Wang, M. Metz and M. Kobrinsky, in *2024 IEEE International Electron Devices Meeting (IEDM)*, 2024, pp. 1–4.
23 E. Cho, W.-J. Son, S. Lee, H.-S. Do, K. Min and D. S. Kim, *J. Mater. Chem. C*, 2025, **13**, 7772–7784.
24 P. Zheng and D. Gall, *Journal of Applied Physics*, 2017, **122**, 135301.
25 K. Sankaran, S. Clima, M. Mees and G. Pourtois, *ECS J. Solid State Sci. Technol.*, 2015, **4**, N3127–N3133.
26 S. Kumar, C. Multunas, B. Defay, D. Gall and R. Sundararaman, *Phys. Rev. Materials*, 2022, **6**, 085002.
27 M. Zhang and C. Adelmann, *Journal of Applied Physics*, 2025, **138**, 090902.
28 G. Kresse and J. Hafner, *Phys. Rev. B*, 1993, **47**, 558–561.
29 G. Kresse and J. Furthmüller, *Phys. Rev. B*, 1996, **54**, 11169–11186.
30 J. P. Perdew, K. Burke and M. Ernzerhof, *Phys. Rev. Lett.*, 1996, **77**, 3865–3868.
31 K. Lejaeghere, G. Bihlmayer, T. Björkman, P. Blaha, S. Blügel, V. Blum, D. Caliste, I. E. Castelli, S. J. Clark, A. Dal Corso, S. De Gironcoli, T. Deutsch, J. K. Dewhurst, I. Di Marco, C. Draxl, M. Dułak, O. Eriksson, J. A. Flores-Livas, K. F. Garrity, L. Genovese, P. Giannozzi, M. Giantomassi, S. Goedecker, X. Gonze, O. Grånäs, E. K. U. Gross, A. Gulans, F. Gygi, D. R. Hamann, P. J. Hasnip, N. A. W. Holzwarth, D. Iuşan, D. B. Jochym, F. Jollet, D. Jones, G. Kresse, K. Koepernik, E. Küçükbenli, Y. O. Kvashnin, I. L. M. Locht, S. Lubeck, M. Marsman, N. Marzari, U. Nitzsche, L. Nordström, T. Ozaki, L. Paulatto, C. J. Pickard, W. Poelmans, M. I. J. Probert, K. Refson, M. Richter, G.-M. Rignanese, S. Saha, M. Scheffler, M. Schlipf, K. Schwarz, S. Sharma, F. Tavazza, P. Thunström, A. Tkatchenko, M. Torrent, D. Vanderbilt, M. J. Van Setten, V. Van Speybroeck, J. M. Wills, J. R. Yates, G.-X. Zhang and S. Cottenier, *Science*, 2016, **351**, aad3000.
32 P. E. Blöchl, *Phys. Rev. B*, 1994, **50**, 17953–17979.
33 A. Dal Corso, *Phys. Rev. B*, 2012, **86**, 085135.
34 A. Jain, S. P. Ong, G. Hautier, W. Chen, W. D. Richards, S. Dacek, S. Cholia, D. Gunter, D. Skinner, G. Ceder and K. A. Persson, *APL Materials*, 2013, **1**, 011002.
35 M. K. Horton, P. Huck, R. X. Yang, J. M. Munro, S. Dwaraknath, A. M. Ganose, R. S. Kingsbury, M. Wen, J. X. Shen, T. S. Mathis, A. D. Kaplan, K. Berket, J. Riebesell, J. George, A. S. Rosen, E. W. C. Spotte-Smith, M. J. McDermott, O. A. Cohen, A. Dunn, M. C. Kuner, G.-M. Rignanese, G. Petretto, D. Waroquiers, S. M. Griffin, J. B. Neaton, D. C. Chrzan, M. Asta, G. Hautier, S. Cholia, G. Ceder, S. P. Ong, A. Jain and K. A. Persson, *Nat. Mater.*, DOI:10.1038/s41563-025-02272-0.
36 W. Sun, S. T. Dacek, S. P. Ong, G. Hautier, A. Jain, W. D. Richards, A. C. Gamst, K. A. Persson and G. Ceder, *Sci. Adv.*, 2016, **2**, e1600225.
37 A. Lee, S. Sarker, J. E. Saal, L. Ward, C. Borg, A. Mehta and C. Wolverton, *Commun Mater*, 2022, **3**, 73.
38 S. P. Ong, W. D. Richards, A. Jain, G. Hautier, M. Kocher, S. Cholia, D. Gunter, V. L. Chevrier, K. A. Persson and G. Ceder, *Computational Materials Science*, 2013, **68**, 314–319.
39 G. K. H. Madsen and D. J. Singh, *Computer Physics Communications*, 2006, **175**, 67–71.
40 G. K. H. Madsen, J. Carrete and M. J. Verstraete, *Computer Physics Communications*, 2018, **231**, 140–145.
41 S. Lim, Y. Kim, G. Maeng and Y. Lee, *J. Phys.: Condens. Matter*, 2026, **38**, 015503.
42 S. Nagata and M. Nakajima, *Physica B: Condensed Matter*, 1993, **192**, 228–232.
43 Z. Hashin and S. Shtrikman, *Phys. Rev.*, 1963, **130**, 129–133.
44 K. Moors, K. Sankaran, G. Pourtois and C. Adelmann, *Phys. Rev. Materials*, 2022, **6**, 123804.
45 S. Park, S. Kang, G. Kim, J. Moon, K. Lee and J. Chang, *IEEE Trans. Electron Devices*, 2025, **72**, 5703–5709.
46 L.-E. Edshammar, H. Iwamoto, G. Bergson, L. Ehrenberg, J. Brunvoll, E. Bunnenberg, C. Djerassi and R. Records, *Acta Chem. Scand.*, 1966, **20**, 427–431.
47 T. D. Boniface and L. A. Cornish, *Journal of Alloys and Compounds*, 1996, **234**, 275–279.
48 W. Jeitschko, H. IIolleck, H. Nowotny and F. Benesovsky, .
49 R. P. Elliott, *Laves phases of the rare earths with transition elements*, Illinois Inst. of Tech., Chicago, IL (United States). IIT Research Inst., 1964.
50 V. N. Eremenko, V. G. Khorujaya, P. S. Martsenyuk and K. Ye. Korniyenko, *Journal of Alloys and Compounds*, 1995, **217**, 213–217.
51 V. G. Khoruzhaya and K. E. Kornienko, *Powder Metallurgy and Metal Ceramics*, 2001, **40**, 362–373.
52 E. Raub, H. Beeskow and W. Fritzsche, *International Journal of Materials Research*, 1963, **54**, 451–454.
53 B. H. Chen and H. F. Franzen, *Journal of the Less Common Metals*, 1990, **157**, 37–45.
54 H. Holleck and H. Kleykamp, *Journal of Nuclear Materials*, 1970, **35**, 158–166.
55 H. R. Ott, F. Hulliger, H. Rudigier and Z. Fisk, *Phys. Rev. B*, 1985, **31**, 1329–1333.
56 E. Raub and W. Fritzsche, *International Journal of Materials Research*, 1963, **54**, 21–23.
57 R. M. Waterstrat and R. C. Manuszewski, *Journal of the Less Common Metals*, 1976, **48**, 151–158.
58 A. Landelli and A. Palenzona, *Revue de Chimie Minerale*, 1976, **13**, 55–61.
59 Y.-Y. Fang, Y.-H. Tsai, Y.-L. Chen, D.-J. Jhan, M.-Y. Lu, P. Y. Keng and S.-Y. Chang, *Applied Physics Letters*, 2024, **124**, 142108.
60 A. A. Menovsky, *Journal of Magnetism and Magnetic Materials*, 1988, **76–77**, 631–636.
61 M. J. Besnus, A. Essaihi, N. Hamdaoui, G. Fischer, J. P. Kappler, A. Meyer, J. Pierre, P. Haen and P. Lejay, *Physica B: Condensed Matter*, 1991, **171**, 350–352.
62 W. Jeitschko, R. Glaum and L. Boonk, *Journal of Solid State Chemistry*, 1987, **69**, 93–100.
63 K. Hiebl, C. Horvath, P. Rogl and M. J. Sienko, *Journal of Magnetism and Magnetic Materials*, 1983, **37**, 287–296.
64 C. Godart, C. Ammarguellat, N. Wetta, G. Krill and J. C. Achard, *Journal of Magnetism and Magnetic Materials*, 1987, **63–64**, 527–528.
65 M. Francois, G. Venturini, J. F. Marêché, B. Malaman and B. Roques, *Journal of the Less Common Metals*, 1985, **113**, 231–237.
66 I. Felner and I. Nowik, *J PHYS CHEM SOLIDS*, 1985, **46**, 681–687.
67 A. A. Menovsky, A. C. Moleman, G. E. Snel, T. J. Gortenmulder, H. J. Tan and T. T. M. Palstra, *Journal of Crystal Growth*, 1986, **79**, 316–321.

68 T. E. Mason, B. D. Gaulin, J. D. Garrett, Z. Tun, W. J. L. Buyers and E. D. Isaacs, *Phys. Rev. Lett.*, 1990, **65**, 3189–3192.
69 B. C. Odegard and C. H. Cadden, in *17th IEEE/NPSS Symposium Fusion Engineering (Cat. No.97CH36131)*, 1997, vol. 2, pp. 896–900 vol.2.
70 N. Sakamoto and K. Hiroshi, *Journal of Nuclear Materials*, 1997, **233–237**, 609–611.
71 L. L. Snead and S. J. Zinkle, *AIP Conf. Proc.*, 2005, **746**, 768–775.
72 O. D. Neĭkov, S. S. Naboychenko and I. B. Murashova, *Handbook of Non-Ferrous Metal Powders*, Elsevier Ltd., Second Edition., 2019.
73 G. Maeng and Y. Lee, *Electron. Mater. Lett.*, DOI:10.1007/s13391-026-00632-9.
74 H. Yamada, C. Liu, S. Wu, Y. Koyama, S. Ju, J. Shiomi, J. Morikawa and R. Yoshida, *ACS Cent. Sci.*, 2019, **5**, 1717–1730.
75 C. Adelmann, L. G. Wen, A. P. Peter, Y. K. Siew, K. Croes, J. Swerts, M. Popovici, K. Sankaran, G. Pourtois, S. Van Elshocht, J. Bömmels and Z. Tőkei, in *IEEE International Interconnect Technology Conference*, 2014, pp. 173–176.
76 K. Croes, Ch. Adelmann, C. J. Wilson, H. Zahedmanesh, O. V. Pedreira, C. Wu, A. Lesniewska, H. Oprins, S. Beyne, I. Ciofi, D. Kocaay, M. Stucchi and Zs. Tokei, in *2018 IEEE International Electron Devices Meeting (IEDM)*, IEEE, San Francisco, CA, 2018, p. 5.3.1-5.3.4.
77 J. M. Luttinger and J. C. Ward, *Phys. Rev.*, 1960, **118**, 1417–1427.
78 J. M. Luttinger, *Phys. Rev.*, 1960, **119**, 1153–1163.
79 S. Kang, S. Park and J. Chang, *J. Mater. Chem. C*, 2026, 10.1039.D5TC04036A.
80 Y. Hu, P. Conlin, Y. Lee, D. Kim and K. Cho, *J. Mater. Chem. C*, 2022, **10**, 5627–5635.
81 J. Koike, M. Haneda, J. Iijima and M. Wada, in *2006 International Interconnect Technology Conference*, IEEE, Burlingame, CA, 2006, pp. 161–163.
82 J. Koike, T. Kuge, L. Chen and M. Yahagi, in *2021 IEEE International Interconnect Technology Conference (IITC)*, IEEE, Kyoto, Japan, 2021, pp. 1–3.
83 C. Kim, G. Kang, Y. Jung, J.-Y. Kim, G.-B. Lee, D. Hong, Y. Lee, S.-G. Hwang, I.-H. Jung and Y.-C. Joo, *Sci Rep*, 2022, **12**, 12291.
84 T. Kuge, M. Yahagi and J. Koike, *Journal of Alloys and Compounds*, 2022, **918**, 165615.
85 M. Zhang and D. Gall, *IEEE Trans. Electron Devices*, 2024, **71**, 3252–3257.
86 M. Iwabuchi, Y. Suto and J. Koike, *Applied Surface Science*, 2026, **720**, 165148.
87 H. Lee, S. Poncé, K. Bushick, S. Hajinazar, J. Lafuente-Bartolome, J. Leveillee, C. Lian, J.-M. Lihm, F. Macheda, H. Mori, H. Paudyal, W. H. Sio, S. Tiwari, M. Zacharias, X. Zhang, N. Bonini, E. Kioupakis, E. R. Margine and F. Giustino, *npj Comput Mater*, 2023, **9**, 156.
88 J.-J. Zhou, J. Park, I.-T. Lu, I. Maliyov, X. Tong and M. Bernardi, *Computer Physics Communications*, 2021, **264**, 107970.

Supplementary Information for

# First-principles high-throughput screening of ruthenium compounds for advanced interconnects

Gyungho Maeng,[a] Subeen Lim,[a] Bonggeun Shong,[b] and Yeonghun Lee*[ac]

[a] *Department of Electronics Engineering, Incheon National University, Yeonsu-gu, Incheon 22012, Republic of Korea.*

[b] *Major in Advanced Materials and Semiconductor Engineering, Hanyang University, Ansan-si, Gyeonggi-do 15588, South Korea.*

[c] *Research Institute for Engineering and Technology, Incheon National University, Yeonsu-gu, Incheon 22012, Republic of Korea.*

* Corresponding author.
*E-mail address:* y.lee@inu.ac.kr *(Y. Lee).*

# 1. Resistivity-based averaging scheme

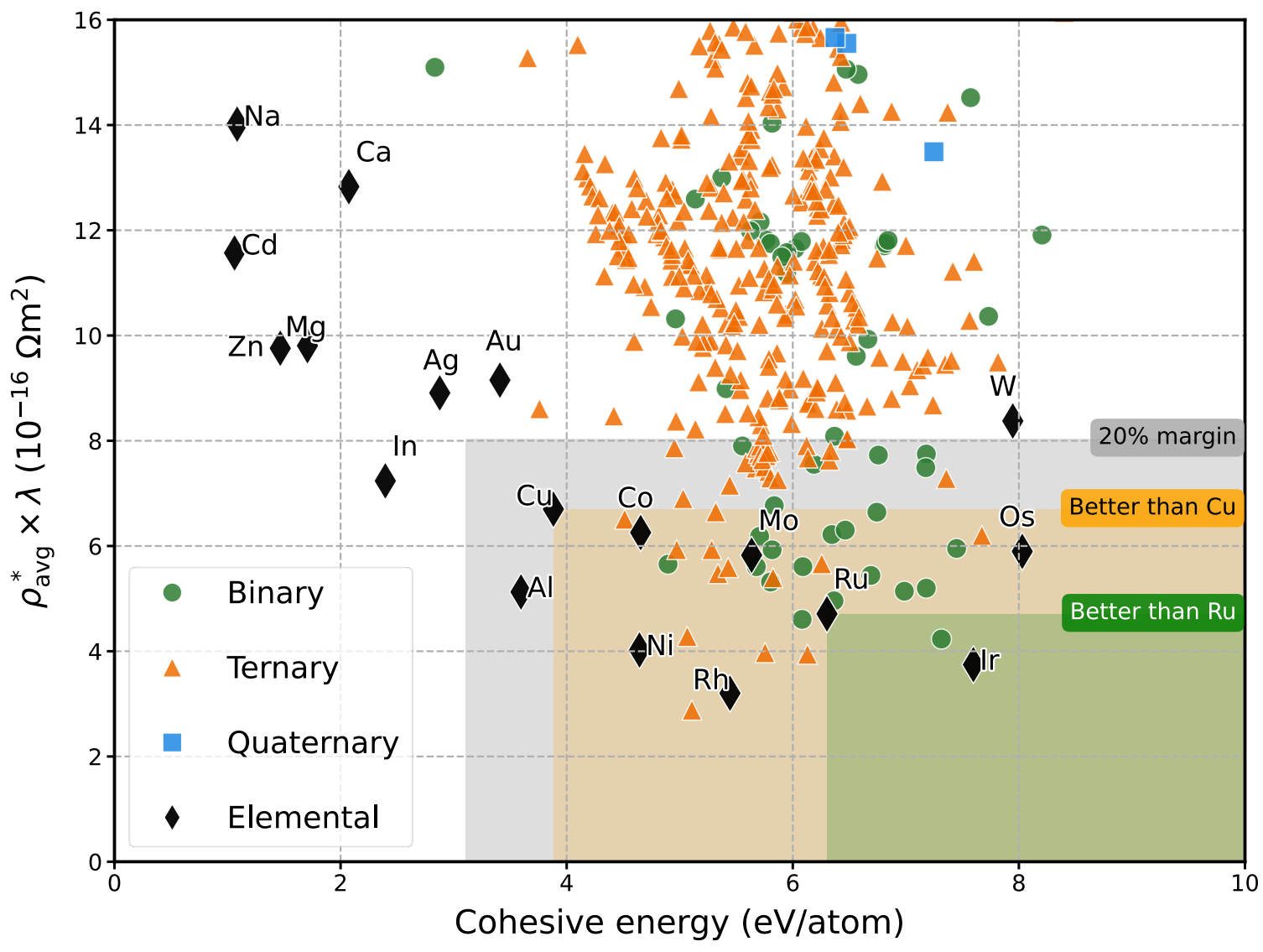

**Fig. S1.** Global screening results using the resistivity-based averaging scheme. The electrical properties ($\rho^*_{avg} \times \lambda$) are plotted as a function of cohesive energy for all screened Ru-based compounds. This plot serves as the resistivity-based counterpart to Fig. 2 in the manuscript. The shaded areas represent the performance margins relative to Cu and Ru. Note that due to the inclusion of high-resistivity direction in the averaging, some anisotropic compounds may exhibit inferior resistivity performance compared to the conductivity-based results.

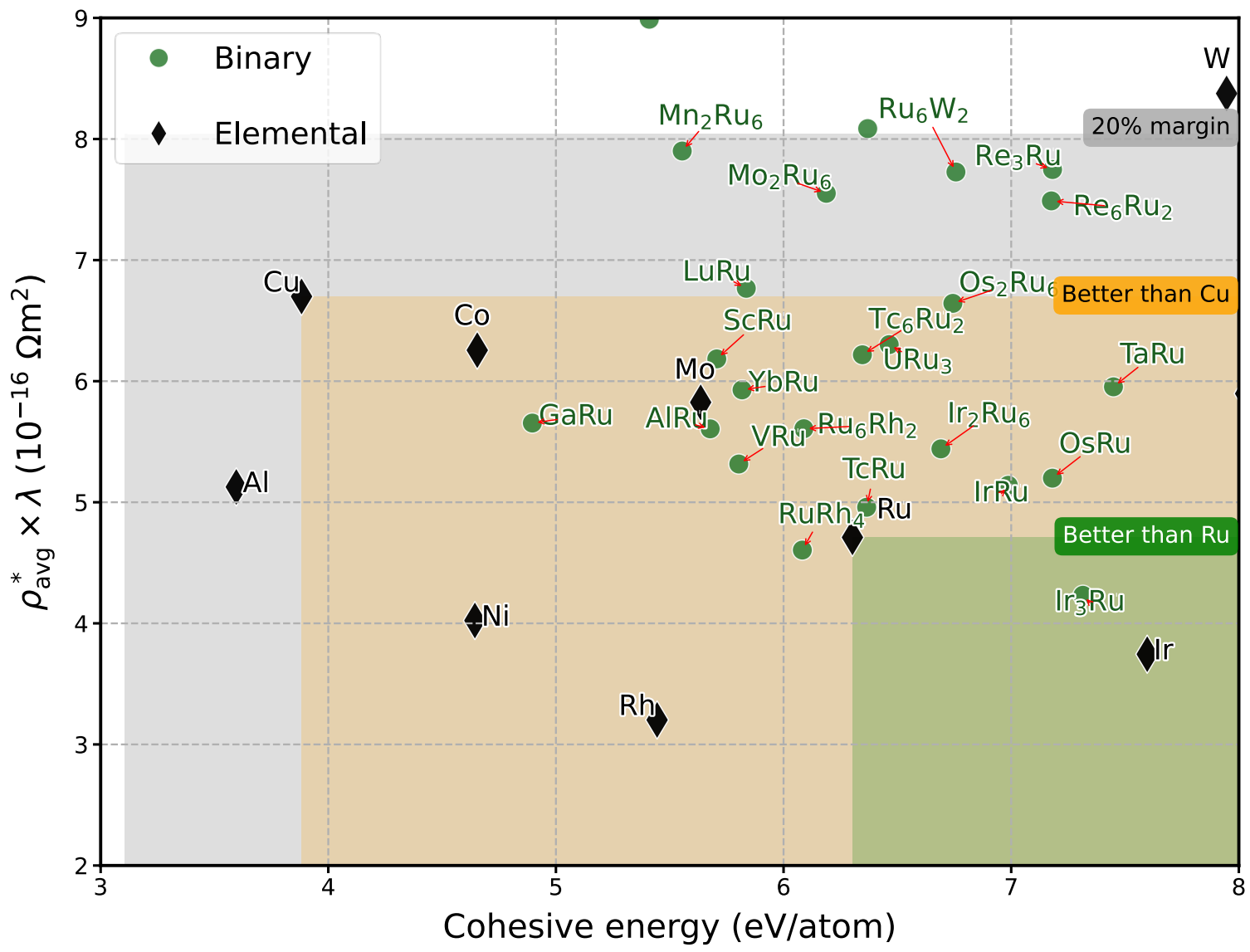

**Fig. S2.** Evaluation of binary Ru compounds based on the resistivity-based metric. The scatter plot displays the distribution of binary candidates against the Cu and Ru benchmarks. This corresponds to Fig. 3 in the manuscript. Notably, the binary compound $Tc_2Ru_6$ falls outside the promising window in this plot due to the increased average resistivity resulting from its anisotropic transport properties. Consequently, under this upper-bound estimation, the number of promising binary candidates is slightly reduced from 23 to 22.

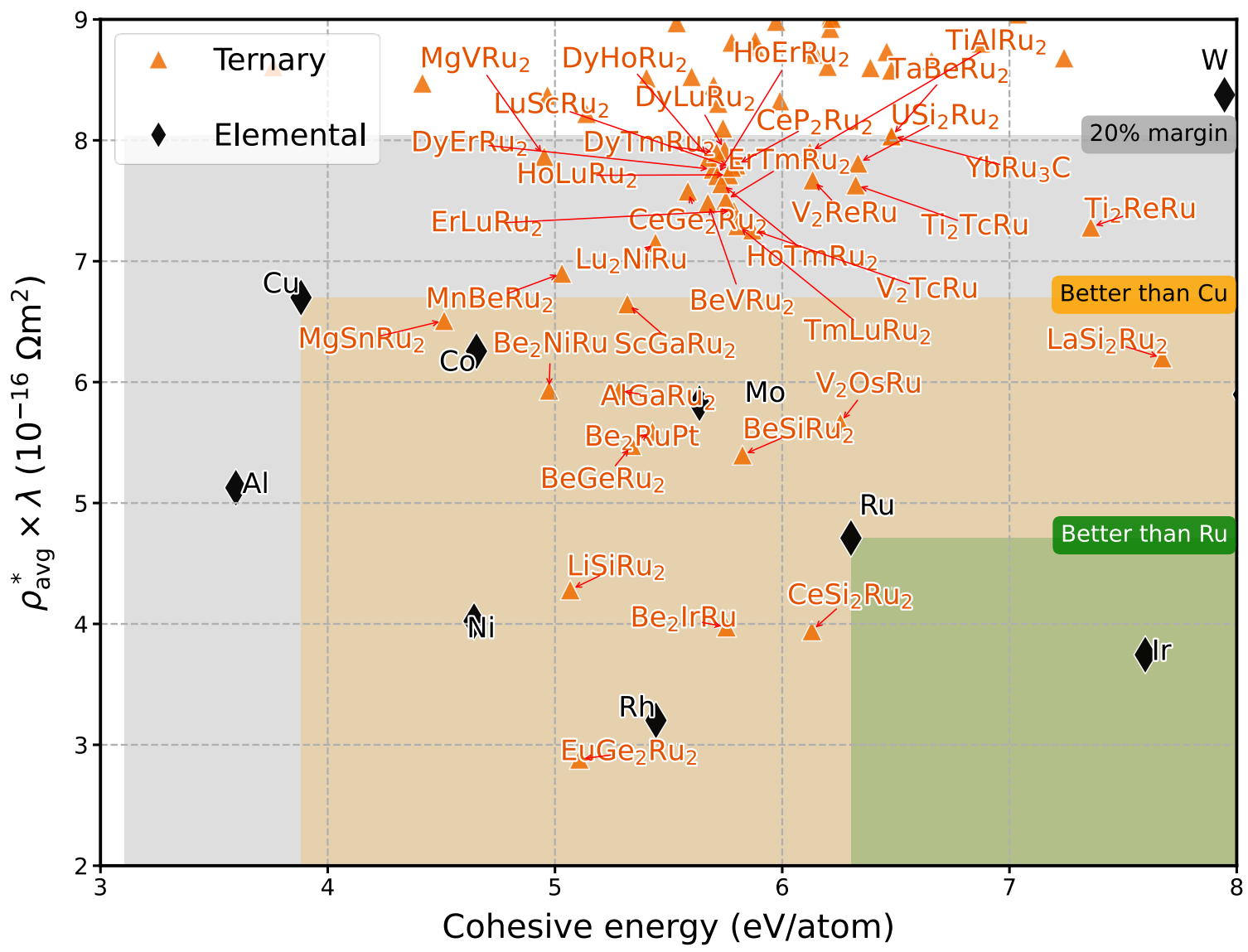

**Fig. S3.** Evaluation of ternary Ru compounds based on the resistivity-based metric. The distribution of ternary candidates is mapped across the performance-stability design space. This corresponds to Fig. 4 in the manuscript. While the total number of promising ternary candidates remains unchanged compared to the conductivity-based scheme, the specific transport properties for six anisotropic compounds are adjusted upward. This shift visually reflects the impact of directional anisotropy when evaluating the theoretical upper limit of resistivity.

As detailed in the manuscript, the primary screening utilized a conductivity-based averaging scheme ($\rho_{avg} \times \lambda$), assuming an effective medium where current flows through multiple crystallographic channels simultaneously. In contrast, this section presents the screening results based on the resistivity-based averaging scheme ($\rho^*_{avg} \times \lambda$), which mimics extreme confinement in narrow interconnects where directional resistances are effectively combined in series. Consequently, this metric highlights the influence of high-resistivity directions and provides a theoretical upper bound on the electrical resistivity. Comparing the results in Figs. S1-S3 with those in the main text (Figs. 2-4), distinct numerical shifts are observed for non-cubic compounds exhibiting strong anisotropy, verifying the performance variation due to directional sensitivity.

## 2. Expanded display of elemental metals

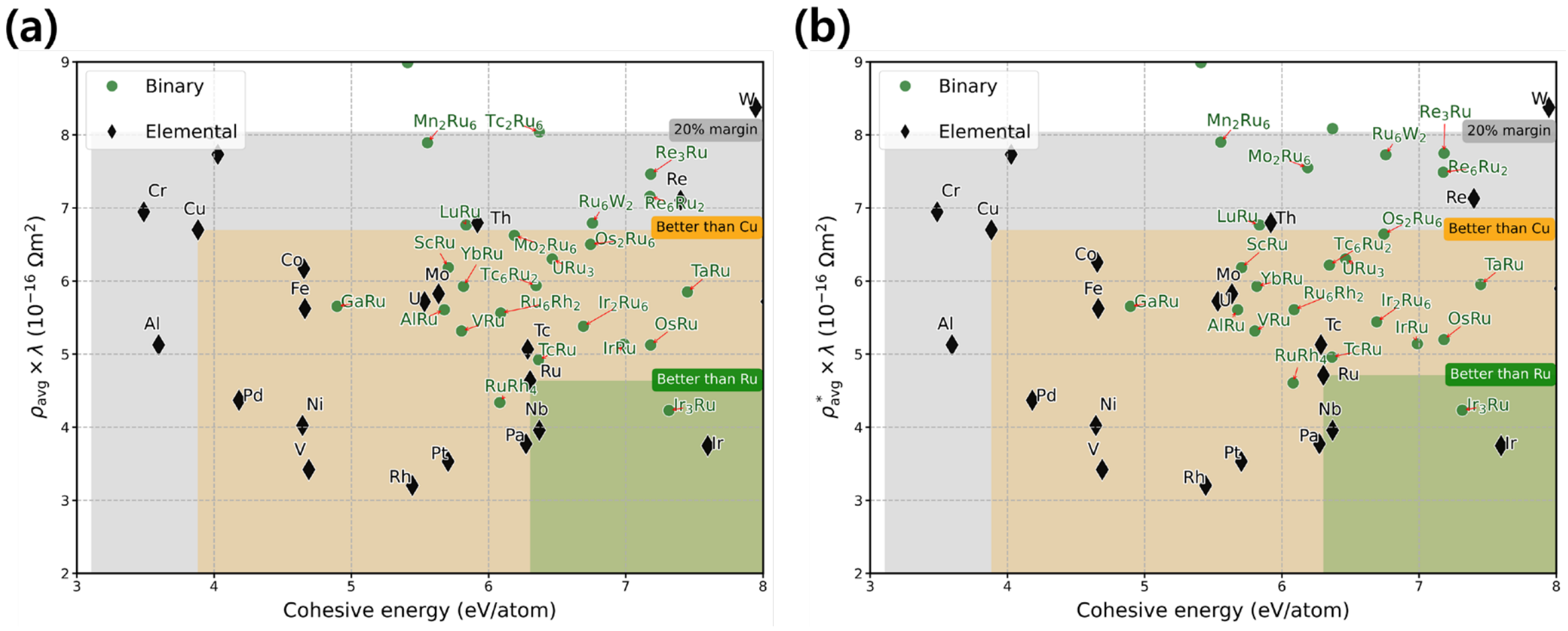

**Fig. S4.** Binary Ru compounds with expanded metallic elements based on the (a) conductivity-based average and (b) resistivity-based average metrics. Elemental metals (black diamonds) are plotted alongside the screened binary compounds (green circles).

The plots display $\rho_{\text{avg}} \times \lambda$ versus cohesive energy ($E_{\text{coh}}$) evaluated via the (a) conductivity-averaged and (b) resistivity-averaged schemes. This expanded analysis visually demonstrates that the performance metrics of a compound typically interpolate between those of its elemental constituents or degrade further. This supports the conclusion that compound formation fundamentally redistributes, rather than inherently transcends, the intrinsic electronic properties of the pure building blocks.

## 3. Compositional descriptors extracted via the XenonPy toolkit

**Table S 1 List of top 10 elements-level compositional features from XenonPy**

| Compositional property | Description |
|---|---|
| vdw_radius_uff | Van der Waals radius from the UFF |
| melting_point | Melting point |
| heat_of_formation | Heat of formation |
| evaporation_heat | Evaporation heat |
| fusion_enthalpy | Fusion heat |
| hhi_r | Herfindahl-Hirschman Index (HHI) reserves values |
| num_s_unfilled | Unfilled electron in s shell |
| num_p_unfilied | Unfilled electron in p shell |
| gs_energy | DFT energy per atom (raw VASP value) of $T = 0$ K ground state |
| gs_est_bcc_latcnt | Estimated BCC lattice parameter based on the DFT volume |
| gs_est_fcc_latcnt | Estimated FCC lattice parameter based on the DFT volume |
| gs_volume_per | DFT volume per atom of $T = 0$ K ground state |
| Bulk_modulus | |

These properties represent the variance of elemental features within each compound and were identified through Pearson correlation analysis as having the strongest relationships with the intrinsic transport metric.